\documentclass[12pt]{iopart}

\usepackage{graphicx}
\expandafter\let\csname equation*\endcsname\relax

\expandafter\let\csname endequation*\endcsname\relax
\usepackage{amsmath}
\usepackage{cite}
\usepackage[colorlinks, citecolor = blue, urlcolor = blue]{hyperref}
\begin{document}

\title[Phonons in stringlet-land and the boson peak]{Phonons in stringlet-land and the boson peak}

\author{Cunyuan Jiang, Matteo Baggioli\footnote{Author to whom any correspondence should be addressed}}

\address{School of Physics and Astronomy, Shanghai Jiao Tong University, Shanghai 200240, China}
\address{Wilczek Quantum Center, School of Physics and Astronomy, Shanghai Jiao Tong University, Shanghai 200240, China}
\address{Shanghai Research Center for Quantum Sciences, Shanghai 201315,China}

\ead{cunyuanjiang@sjtu.edu.cn, b.matteo@sjtu.edu.cn}
\vspace{10pt}
\begin{indented}
\item[]\today
\end{indented}

\begin{abstract}
Solid materials that deviate from the harmonic crystal paradigm exhibit characteristic anomalies in the specific heat and vibrational density of states (VDOS) with respect to Debye's theory predictions. The boson peak (BP), a low-frequency excess in the VDOS over Debye law $g(\omega) \propto \omega^2$, is certainly the most famous among them; nevertheless, its origin is still subject of fierce debate. Recent simulation works provided strong evidence that localized one-dimensional string-like excitations (stringlets) might be the microscopic origin of the BP. In this work, we study the dynamics of acoustic phonons interacting with a bath of vibrating 1D stringlets with exponentially distributed size, as observed in simulations. We show that stringlets strongly renormalize the phonon propagator and naturally induce a BP anomaly in the vibrational density of states, corresponding to the emergence of a dispersionless BP flat mode. Additionally, phonon-stringlet interactions produce a strong enhancement of sound attenuation and a dip in the speed of sound near the BP frequency, consistent with experimental and simulation data. The qualitative trends of the BP frequency and intensity are predicted within the model and shown to be in good agreement with previous observations. In summary, our results substantiate with a simple theoretical model the recent simulation results by Hu and Tanaka claiming the origin of the BP from stringlet dynamics.
\end{abstract}

%
%
%
%
%

\section{Introduction}
The low-energy properties of crystalline solids with long-range order can be rationalized within the Debye's paradigm \cite{debye}, and explained by the harmonic dynamics of acoustic propagating phonon modes, the Goldstone modes of translational symmetry \cite{Leutwyler:1996er}. The most fundamental result of Debye's theory is that the low-frequency vibrational density of states (VDOS) of ideal crystals follows a power law $g(\omega) \propto \omega^{d-1}$ \cite{kittel2021introduction}, where $d$ is the number of spatial dimensions. 

Notably, amorphous solid materials \cite{ramos2022low}, but interestingly also several crystalline compounds with minimal or absent structural disorder \cite{PhysRevB.23.3886,RevModPhys.86.669,Schliesser_2015,PhysRevB.99.024301}, defy this paradigm. For three-dimensional amorphous systems, the VDOS reduced by Debye's power law, $g(\omega)/\omega^2$, is not a constant at low frequencies but it rather shows a pronounced peak which is known as the ``boson peak'' (BP). Despite the uncountable simulation and experimental observations of this phenomenon, which are impossible to be all listed here (see \cite{ramos2022low} for a recent book on the topic), a general consensus on its microscopic origin has not been achieved yet and several theoretical models have been proposed in the past, \textit{e.g.}, \cite{Elliott_1992,Schirmacher_2006,PhysRevLett.98.025501,PhysRevLett.115.015901,PhysRevLett.97.055501,PhysRevB.43.5039,PhysRevB.67.094203,PhysRevB.76.064206,PhysRevE.61.587,PhysRevLett.122.145501,PhysRevLett.86.1255,PhysRevLett.106.225501,doi:10.1080/00018738900101162,KLINGER2002311,Grigera2003,jiang2023stringlet}.

\begin{figure}[ht]
    \centering
    \includegraphics[width=0.6\linewidth]{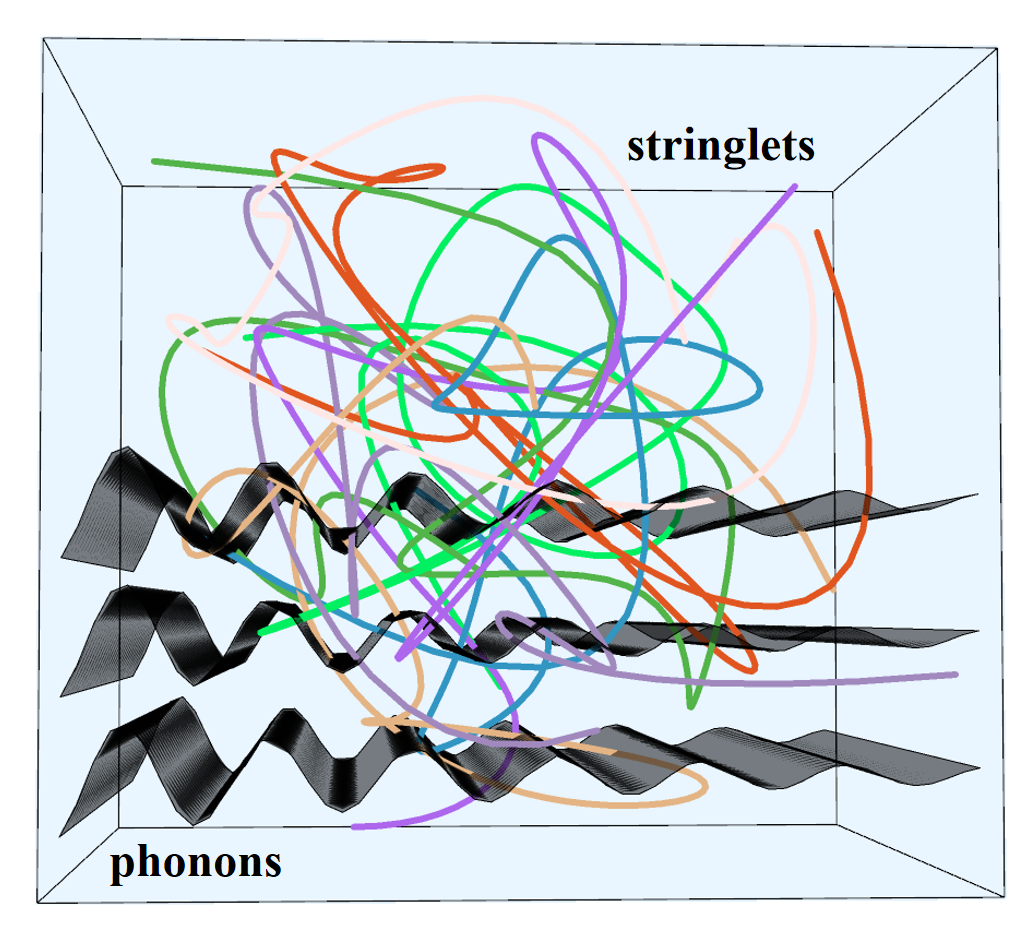}
    \caption{\textbf{Phonon dynamics in stringlets-land.} Acoustic waves interacting with localized 1D string-like objects with exponentially distributed size -- ``stringlets''.}
    \label{fig0}
\end{figure}

Among the various frameworks, a common interpretation is based on the presence of quasi-localized modes which coexist with phonons \cite{PhysRevB.53.11469,10.1063/5.0147889,10.21468/SciPostPhysCore.4.2.008,2348f3ae65af4aef9f03dc50fda47415,SCHOBER1993965}, and that are possibly, but not necessarily, emerging because of the underlying structural disorder.

Along these lines, recent simulation results in $2$D and $3$D model glasses \cite{PhysRevResearch.5.023055,Hu2022}, and in heated glass-forming liquids as well \cite{d1,C2SM26789F,C2SM27533C,Douglas_2016,d3}, have revealed the crucial role of localized one-dimensional string-like excitations, termed ``stringlets", that have been ascribed as the microscopic origin of the BP (see Fig.\ref{fig0} for a cartoon of stringlets coexisting with phonons in amorphous solids). 

Importantly, there is experimental evidence from low-frequency Raman scattering data that the excess modes responsible for the BP are predominantly of transverse nature and one-dimensional \cite{PhysRevB.59.38,PhysRevB.48.12539}, consistent with the stringlet description. The idea that dislocation-like defects might have an important function in glasses is not new \cite{novikov1990spectrum,Angell_2004,BHAT20064517,RevModPhys.80.61}, but it has recently re-emerged in the discussion regarding the origin of the BP and also in the challenge of identifying the plastic mediators and ``soft spots" in amorphous solids \cite{Cao2018,PhysRevLett.127.015501,Baggioli2023,PhysRevE.105.024602,Wu2023,desmarchelier2024topological,bera2024soft}.

Inspired by the original elastic string model by Granato and L\"{u}cke \cite{granato1956theory}, Lund and collaborators have extensively studied the problem of the interplay between elastic phonons and dislocation-like line defects \cite{PhysRevB.70.024303,PhysRevB.72.174110,PhysRevB.72.174111,PhysRevB.91.094102,PhysRevB.106.024105,Lund1988}, and even pushed it forward as a possible explanation for glassy anomalies \cite{PhysRevB.101.174311}. In its simplest incarnation, the theory models the defects as a collection of one-dimensional vibrating lines pinned at their extremes and whose length $l$ is distributed according to an unknown function $p(l)$. At the time of Ref. \cite{PhysRevB.101.174311}, no simulation data for $p(l)$ were available; therefore, Lund et al. \cite{PhysRevB.101.174311} through reverse engineering used experimental data for glycerol and silica to extract $p(l)$ and found that a gaussian-shaped distribution would give impressive agreement with the data.

Unfortunately, as we now know from various works \cite{PhysRevResearch.5.023055,Hu2022,d1,C2SM26789F,C2SM27533C,Douglas_2016,d3}, a gaussian-shaped distribution is not compatible with the simulation data that, on the contrary, show a monotonically decreasing exponential distribution of lengths $p(l)\sim \exp(-l/\lambda)$, with possible power-law corrections \cite{10.1063/1.4878502}. Additionally, without the direct input of $p(l)$ from simulations, the theory in Ref. \cite{PhysRevB.101.174311} cannot be fully predictive since $p(l)$ has to be fitted from the data themselves. In \cite{jiang2023stringlet}, we have revisited Lund's model \cite{PhysRevB.101.174311} by assuming the correct exponential distribution for the stringlet length and we have found remarkable quantitative agreement with the simulation data of \cite{PhysRevResearch.5.023055,Hu2022} using a parameter-free theoretical formula for the BP frequency in the low-temperature glassy phase.

Nonetheless, in \cite{jiang2023stringlet}, for simplicity, we have considered phonons and stringlets as independent entities and derived the low-frequency excess anomaly uniquely considering the contribution to the vibrational density of states from the stringlets. Technically, this assumption implies that the total vibrational density of states is just the sum of a Debye phononic contribution and an independent term arising from the stringlet modes. This decoupling of the modes seems to break down at large frequencies where hybridization takes place (see, \textit{e.g.}, \cite{10.1063/5.0147889}), and it is possibly the responsible for the large uncertainties found when compared to the data at large temperatures, well above the phenomenological glass transition temperature $T_g$ (see \cite{jiang2023stringlet} for details).

\begin{figure*}
    \centering
    \includegraphics[width=\linewidth]{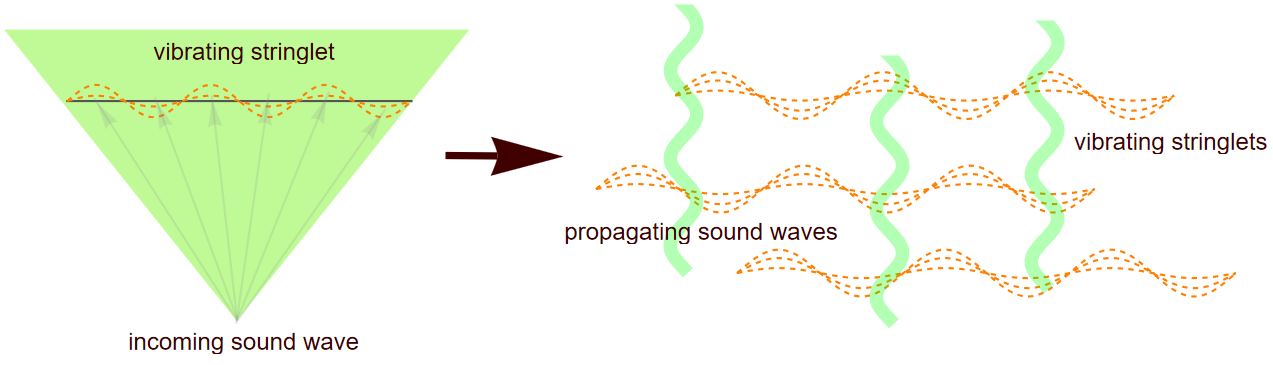}
    \caption{\textbf{A representation of the theoretical model for stringlet-phonon interactions.} An incoming sound wave excites a stringlet that starts vibrating around its equilibrium position. Consequently, sound waves propagate in a bath of vibrating stringlets that strongly renormalize their properties as described by Eq.\eqref{g1}.}
    \label{fig:1}
\end{figure*}

In this work, we extend the theoretical model presented in \cite{jiang2023stringlet} by considering the interplay between stringlets and phonons. Among the various advantages of this route, we are now able to predict the effects of the stringlets on the phonon properties such as their propagation speed and their attenuation constant.

Before continuing, we emphasize that within our model we consider string-like vibrations as the only additional low-energy modes that coexist with acoustic phonons in glasses. Despite other non-phononic modes likely exist in glasses, \textit{e.g.}, quadrupolar four-leaf defects \cite{10.1063/5.0069477}, several recent simulation results \cite{Hu2022,PhysRevResearch.5.023055,10.1063/5.0197386} have proved that they are not fundamental for the BP. Our assumption is therefore supported by simulation results; possible extensions beyond this simplification will be considered in future research.
\section{Theoretical model}
The theoretical model considered in this work was introduced by Lund and collaborators in a series of works \cite{PhysRevB.70.024303,PhysRevB.72.174110,PhysRevB.72.174111,PhysRevB.91.094102,PhysRevB.106.024105,Lund1988,PhysRevB.101.174311}. For simplicity, we will use a simplified version of Lund's model that does not take into account the different phonon polarizations (transverse and longitudinal) and other complications related to the properties of the dislocation-like line defects. In a sense, our model is more similar to the scalar elasticity theory proposed by Granato and L\"{u}cke \cite{granato1956theory}. This simplifying assumption is motivated by several simulation and experimental works that proved that longitudinal modes are not essential to understand glassy anomalies and the physics of the boson peak. It is indeed known that the vibrational density of states and the position and height of the boson peak are dominated by the contribution of transverse modes (see for example Fig.1 in \cite{PhysRevResearch.5.023055}). We find therefore reasonable and justified to simplify the framework by considering a single phononic polarization. We emphasize that this hypothesis does not correspond to describe a one-dimensional solid but rather to consider a three-dimensional solid where only one phonon polarization is active.

In what follows, we will consider the computation of the phonon self-energy only at leading order in the potential induced by the interactions with the stringlets. As we will explicitly prove, this simplified model is enough to capture the salient physical features. At the same time, its generalization is straightforward and indeed already built by Lund and collaborators in several works \cite{PhysRevB.72.174110,Lund1988,PhysRevB.91.094102,PhysRevB.106.024105,PhysRevB.72.174111,PhysRevB.70.024303}.  Finally, before starting our exposition, we emphasize that the main difference between our analysis on glassy anomalies and the one in \cite{PhysRevB.101.174311} relies on the distribution of the stringlet lengths $p(l)$. In \cite{PhysRevB.101.174311}, the distribution $p(l)$ was not taken as an input of the theory, but rather fitted from the data, leading to an incorrect final result. On the other hand, it is now derived explicitly thanks to the recent simulation progress \cite{PhysRevResearch.5.023055,Hu2022,d1,C2SM26789F,C2SM27533C,Douglas_2016,d3} that allow for a direct determination of $p(l)$. This also strongly improves the degree of predictability of the theoretical model, as already confirmed in \cite{jiang2023stringlet}.

The main idea of this theoretical description, visualized in Fig.\ref{fig:1}, is that the elastic medium contains one-dimensional string-like defects (stringlets), pinned at their endpoints, that are excited by incoming sound waves and vibrate as a consequence of that. As in a chain reaction, sound waves in the medium now propagate in a bath of vibrating one-dimensional defects that strongly renormalize sound propagation, and the properties of phonons, with crucial consequent effects for the vibrational dynamics of the medium.

As explained in \cite{PhysRevB.101.174311}, this is equivalent to considering the propagation of sound in a medium with a frequency dependent index of refraction induced in this concrete case by the presence of the stringlets. Neglecting the different longitudinal and transverse polarizations (see Appendix A in \cite{PhysRevB.101.174311} for a more complete treatment), the phonon Green function can be written as
\begin{equation}
    \mathcal{G}(k,\omega)=\frac{1}{-\omega^2 +v^2 k^2-i \omega \gamma k^2- \Gamma(\omega,k)},\label{g1}
\end{equation}
where $\Gamma(\omega,k)$ is the self-energy including the effects coming from the stringlets and $v,\gamma$ are respectively the bare speed of sound and phonon damping. Additionally, $\omega,k$ are the frequency and wave-vector of the phonon. Eq.\eqref{g1} is the result of a Dyson's equation,
\begin{equation}\label{dys}
    \mathcal{G}(\omega,k)= \left[\mathcal{G}_0(\omega,k)^{-1}-\Gamma(\omega,k)\right]^{-1},
\end{equation}
where $\mathcal{G}_0(\omega,k)$ is the ``bare'' phonon Green's function
\begin{equation}\label{bareph}
    \mathcal{G}_0(k,\omega)=\frac{1}{-\omega^2 +v^2 k^2-i \omega \gamma k^2}.
\end{equation}
Here, with the label ``bare'' we mean the phonon propagator before taking into account stringlet-phonon interactions, but taking into account other sources of damping, \textit{e.g.}, phonon-phonon interactions giving rise to Akhiezer damping \cite{Akhiezer}.

By assuming that the length of the stringlets follow a statistical distribution $p(l)$, and working at leading order in the stringlet potential, one obtains the following expression for the self-energy,
\begin{equation}
    \Gamma(\omega,k)  =\mu^2 k^2 \int_0^\infty l\,p(l)\,\mathcal{G}_s(\omega,l)\, dl,\label{selfie}
\end{equation}
where $\mathcal{G}_s(k,\omega)$ is the bare stringlet Green's function:
\begin{equation}\label{vuvu}
    \mathcal{G}_s(\omega,l)=\frac{1}{-\omega^2 +v_s^2 (\pi/l)^2-i \omega \gamma_s}.
\end{equation}
Here we have assumed, as in \cite{jiang2023stringlet}, that the stringlets vibrate only at their fundamental frequency, $\omega^{(l)}_s\equiv \pi v/l$, and neglected all higher harmonics. In Eq.\eqref{vuvu}, $v_s$ and $\gamma_s$ are the stringlet propagation speed and attenuation constant that, for the moment, are taken as independent parameters. Moreover, $\mu$ is a phenomenological parameter with units $\left[\sqrt{L}/t^2\right]$ that parametrizes the strength of the interactions between stringlets and phonons, and that can be eventually related to microscopic physics observables \cite{PhysRevB.101.174311}. For simplicity, in this work, we will take it as an adjustable parameter. To avoid clutter, we do not repeat the derivation of Eq.\eqref{selfie} here but refer the Readers to Appendix \ref{details} for more details. We notice that our equation \eqref{selfie} can be directly compared with Eq.(A.14) in \cite{PhysRevB.101.174311} by identifying $f(\omega)k^2 \equiv \Gamma(\omega,k)$, $v^2 f_0\equiv \mu^2$, and by neglecting the term proportional to $\omega^3$ in the denominator of Eq.(A.14) that comes from second order perturbation theory. 

Having the renormalized Green's function, several other properties of phonons in ``stringlet-land'' follow. More precisely, the dynamic structure factor can be then obtained as,
\begin{equation}
    S(\omega,k)=\dfrac{1}{\pi} \mathrm{Im}[\mathcal{G} (\omega,k)].\label{dsf}
\end{equation}
The phonon vibrational density of states (VDOS) is then given by,
\begin{equation}
    g(\omega)=\dfrac{2 \omega}{k_D^3 \pi} \int_0^{k_D} \mathrm{Im}[\mathcal{G} (\omega,k)] k^2 dk,\label{totaldos}
\end{equation}
where $k_D$ is the Debye wave-vector.
Finally, the real and imaginary parts of the self-energy $\Gamma(\omega,k)$ renormalize the bare speed of sound $v$ and the bare attenuation constant $\gamma$ respectively. Even more drastically, the renormalized Green's function in Eq.\eqref{g1} will contain a more complex pole structures than its bare counterpart, leading to the presence of a flat BP mode.

Before continuing with the results, let us establish our main assumption that is an exponential distribution of the stringlet lengths given by:
\begin{equation}
    p(l)=p_0 e^{-l/\lambda},\label{main}
\end{equation}
where $p_0$ is a normalization factor and $\lambda$ relates to the average stringlet length $\langle l \rangle$. We notice that Eq.\eqref{main} is consistent with several simulation results \cite{PhysRevResearch.5.023055,Hu2022,d1,C2SM26789F,C2SM27533C,Douglas_2016,d3}, and has already been successfully used in the prediction of the BP frequency in the low-temperature glassy state \cite{jiang2023stringlet}. Let us clarify that the stringlet length $l$ does not vary in the interval $[0,\infty]$ but presents a natural IR cutoff in the sample size $L$, and a natural UV cutoff in the atomic distance $a$. As a consequence, the stringlet size distribution in Eq.\eqref{main} has to be limited within the range $a \ll l \ll L$. However, in realistic situations, the average stringlet length, that will set the boson peak scale, is of the order of nanometers, far away from both cutoffs. This implies that the effects of the cutoffs become important only for frequencies much smaller and much larger than the BP frequency, and are therefore irrelevant for the present discussion. 
\begin{figure}
    \centering
    \includegraphics[width=0.48\linewidth]{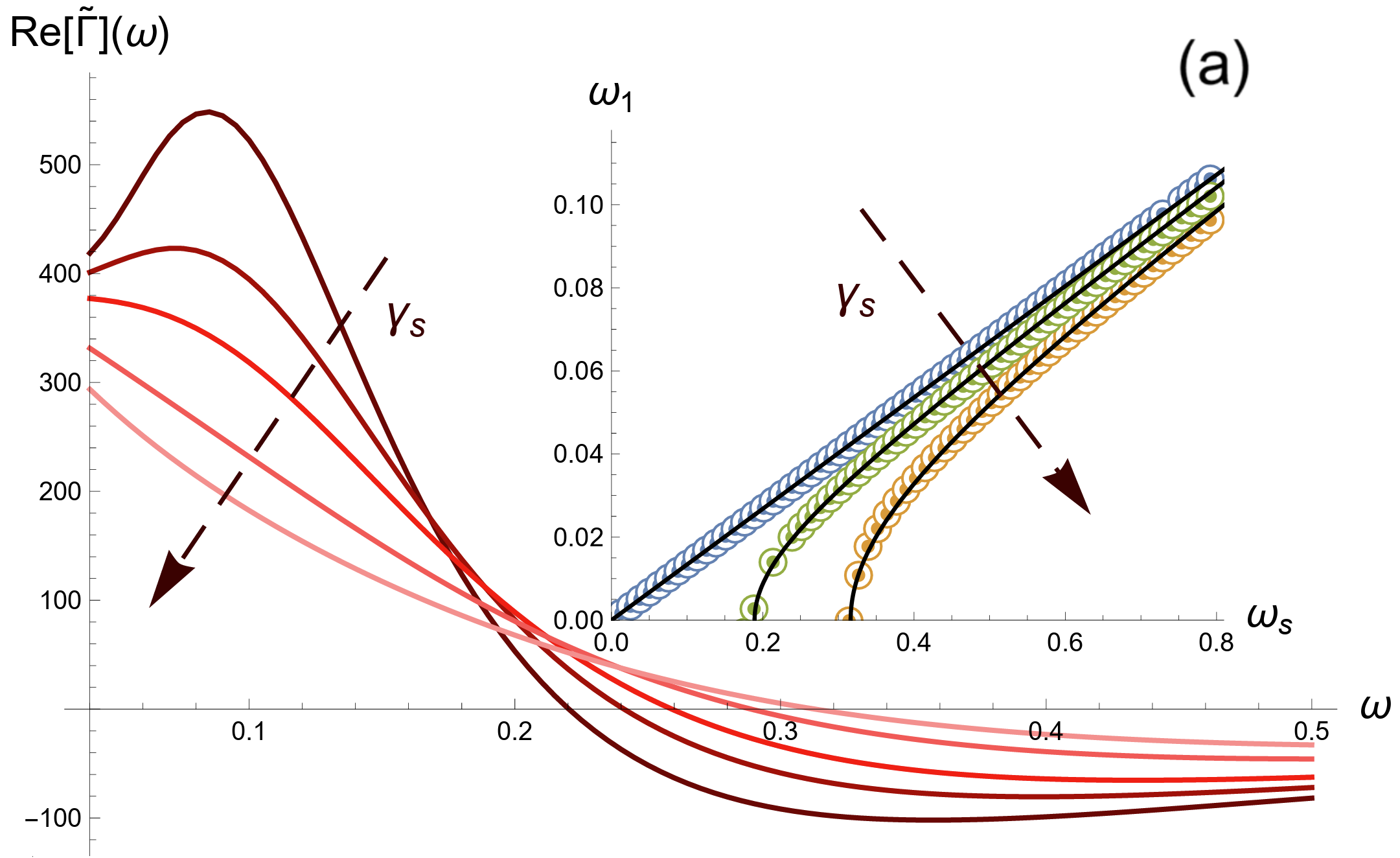} \quad
    \includegraphics[width=0.48\linewidth]{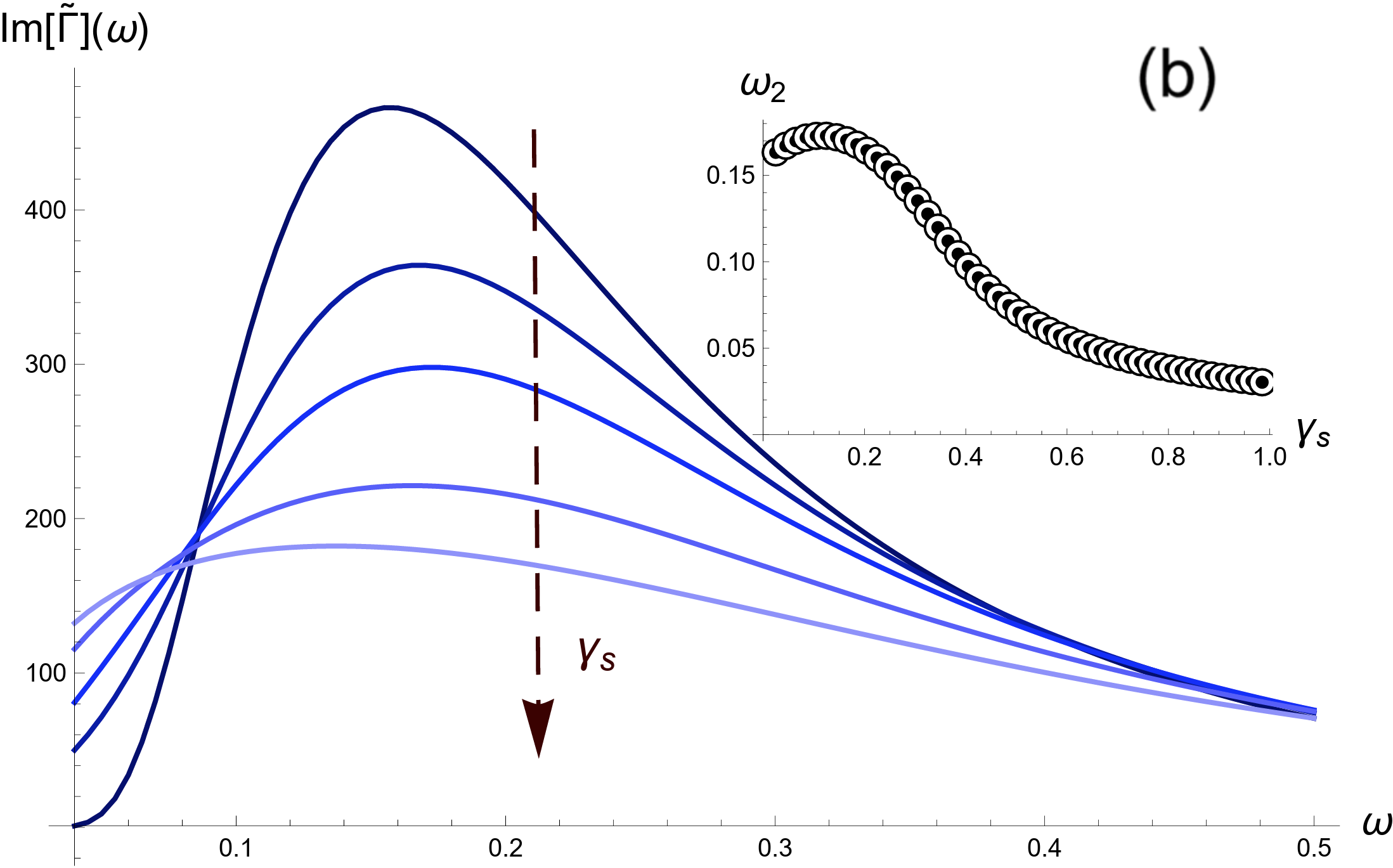}
    
    \caption{\textbf{The phonon self energy.} \textbf{(a)} The real part of the reduced self energy $\tilde \Gamma(\omega)$ as a function of the frequency for $v_s=1$, $\lambda=5$ and different values of $\gamma_s=0,0.05,0.1,0.2,0.3$ from dark to light red. The inset shows the frequency of the maximum $\omega_1$ as a function of the average stringlet frequency $\omega_s$ for $\gamma_s=0,0.03,0.05$. The black lines are the fits discussed in the main text. \textbf{(b)} The imaginary part of $\tilde \Gamma(\omega)$ as a function of $\omega$ for the same values of $\gamma_s$ as in the top panel. The inset shows the frequency of the maximum $\omega_2$ as a function of the stringlet damping $\gamma_s$.}
    \label{fig:2}
\end{figure}

Before moving to our results, let us discuss in detail the various parameters entering in the theoretical model. At first sight, our model contains the following parameters: the phonon and stringlet damping parameters $\gamma,\gamma_s$, the sound and stringlet propagation speeds $v,v_s$, the stringlet-distribution parameters $p_0,\lambda$, the phonon-stringlet coupling $\mu$ and finally the Debye wave-vector $k_D$ (or equivalently the Debye frequency $\omega_D$). In low-temperature glasses, the damping parameters can be neglected as anharmonicities are strongly suppressed. Therefore, we can safely set $\gamma,\gamma_s \rightarrow 0$, as it will be done in the rest of this manuscript. Furthermore, the value of the Debye cutoff does not play any role in our analysis since it corresponds to energies that are much higher than the scales of interest. At the same time, the normalization of the stringlet distribution $p_0$ affects only the intensity of the BP but has no influence on any of the qualitative features discussed in the following, and in particular on the position of the BP.

Finally, in order to reduce the number of free parameters, we set the stringlet propagation speed $v_s$ to be equal to the speed of transverse sound $v$. This working hypothesis can be motivated by previous observations in simulations (\textit{e.g.}, \cite{Shintani2008}) and measurements \cite{PhysRevLett.69.1540,PhysRevLett.78.2405} that the BP strongly correlates with transverse sound modes. This hypothesis has been also verified directly against simulation data in \cite{jiang2023stringlet}.

All in all, we are left with three independent parameters: $v,\lambda,\mu$. We notice that both $v$ and $\lambda$ are measurable and can be directly extracted from the simulation data. See for example Table I in \cite{PhysRevResearch.5.023055} where the values of these two quantities are tabulated for various models. The only left parameter that cannot be determined by simulations and experiments is the stringlet-phonon coupling $\mu$ that is therefore taken as an adjustable parameter in our model. A more microscopic model of stringlet-phonon interactions might provide more information about this coupling.

\section{Results}
\subsection{Phonon self-energy}
Our theoretical analysis starts with a detailed investigation of the phonon self-energy $\Gamma(\omega,k)$. To limit the number of unknown parameters, we neglect the effects of the phonon damping $\gamma$, and we set it to zero if not indicated otherwise. Moreover, to simplify the presentation, we define a renormalized self-energy
\begin{equation}
    \tilde \Gamma (\omega)\equiv \frac{\Gamma(\omega,k)}{\mu^2 k^2},
\end{equation}
which now depends only on the frequency $\omega$ and it is independent of the coupling strength $\mu$. The real and imaginary parts of this renormalized self-energy are plotted as a function of frequency in Fig.\ref{fig:2} for various values of the stringlet damping parameter $\gamma_s$.  For further discussion, we define with
\begin{equation}
    \omega_1 \equiv \max_{\{\omega\}} \mathrm{Re}\tilde \Gamma(\omega),\qquad \omega_2 \equiv \max_{\{\omega\}} \mathrm{Im}\tilde \Gamma(\omega),
\end{equation}
the values of the frequency at which the real and imaginary parts of the renormalized self-energy attain their maximum. The imaginary part of $\tilde \Gamma$ displays a peak that, in the limit of zero stringlet damping, is analytically given by the following expression
\begin{equation}
    \omega_2(\gamma_s=0)=\frac{\pi v_s}{4 \lambda}=\frac{\omega_s}{4}.
\end{equation}
Interestingly, this expression coincides exactly with the theoretical prediction for the BP frequency in the stringlet model of \cite{jiang2023stringlet}, which resulted in good agreement with the simulation data at low temperature (where the damping mechanisms can be neglected). This peak is consistently broadened by increasing the stringlet damping and its position follows a non-monotonic function shown in the inset of the bottom panel of Fig.\ref{fig:2}.

The behavior of the real part of the renormalized self-energy $\tilde \Gamma$ as a function of frequency is more complex. Its trend is characterized by a constant zero frequency value, a possible intermediate peak and high-frequency regime in which the function takes negative values. In the limit of zero stringlet damping, the intermediate peak frequency $\omega_1$ is a linear function of the average stringlet frequency $\omega_s\equiv v \pi/\lambda$. More in general, we find that the intermediate peak frequency in first approximation follows
\begin{equation}\label{eqo}
    \omega_1\approx \sqrt{\omega_s^2-\zeta},
\end{equation}
where $\zeta$ is a phenomenological parameter that vanishes in absence of stringlet damping, and grows with the latter. Eq.\eqref{eqo} provides a good fit to the numerical data presented in the inset of panel (a) in Fig.\ref{fig:2}. Physically, the phenomenological parameter $\zeta$ captures the softening of the frequency scale $\omega_1$ with respect to the bare frequency $\omega_s$ that is defined in absence of damping. We notice that this functional form implies a crossover from an underdamped regime in which the real part of $\tilde \Gamma$ shows a distinct and sharp maximum, to an overdamped regime in which it decays monotonically with the frequency $\omega$.

\subsection{Dynamic structure factor}
Starting from the self-energy, we can directly derive the phonon Green function, Eq.\eqref{g1}, and consequently the dynamic structure factor, Eq.\eqref{dsf}.

\begin{figure}
    \centering
    \includegraphics[width=\linewidth]{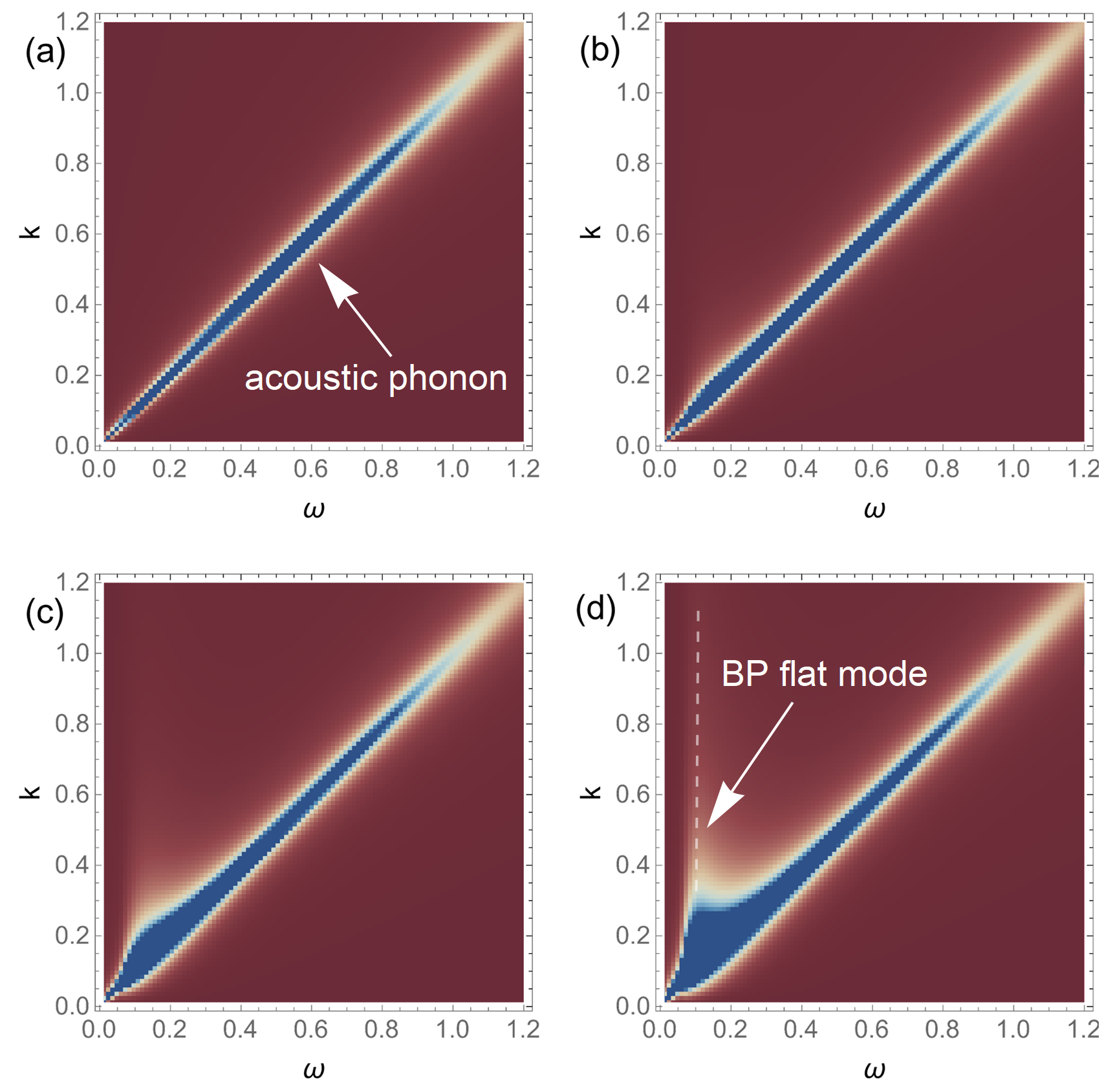}%
    \caption{\textbf{Emergence of a BP flat mode.} The dynamical structure factor $S(\omega,k)$ for $\mu=0,0.01,0.02,0.03$ from \textbf{(a)} to \textbf{(d)}. From red to blue the intensity is stronger. Other parameters involved here are $v=v_s = 1$, $\lambda=5$, $\gamma=\gamma_s=10^{-8}$. The vertical dashed line in panel (d) highlights the flat dispersion of the BP mode that is confirmed by constant $k$ cuts of $S(\omega,k)$ shown in Fig.\ref{figs2}.}
    \label{figdsf}
\end{figure}

If we neglect the interactions between the stringlet degrees of freedom and the acoustic phonons, \textit{i.e.}, $\mu=0$, the dynamic structure factor simply reads
\begin{equation}
    S(\omega,k)_{\mu=0}=\frac{\omega\,\gamma\,k^2}{\pi\,\left[\left(\omega^2-v^2 k^2\right)^2+ \gamma^2 \omega^2 \,k^4\right]}.
\end{equation}
This case is shown in panel (a) of Fig.\ref{figdsf}. The linear dispersion of the damped acoustic phonon is evident. 

By increasing the interaction strength $\mu$, the stringlets start dressing the phonon propagator as progressively shown moving from panel (a) to panel (d) in Fig.\ref{figdsf}. For large enough coupling, panel (d) in Fig.\ref{figdsf}, a dispersionless flat mode emerges below the phonon frequency. As we will see in the next section, the energy of this flat mode coincides with the BP frequency in the reduced density of states. Because of this reason, we label this mode ``the BP mode".

\begin{figure}[htb]
    \centering
    \includegraphics[width=0.7\linewidth]{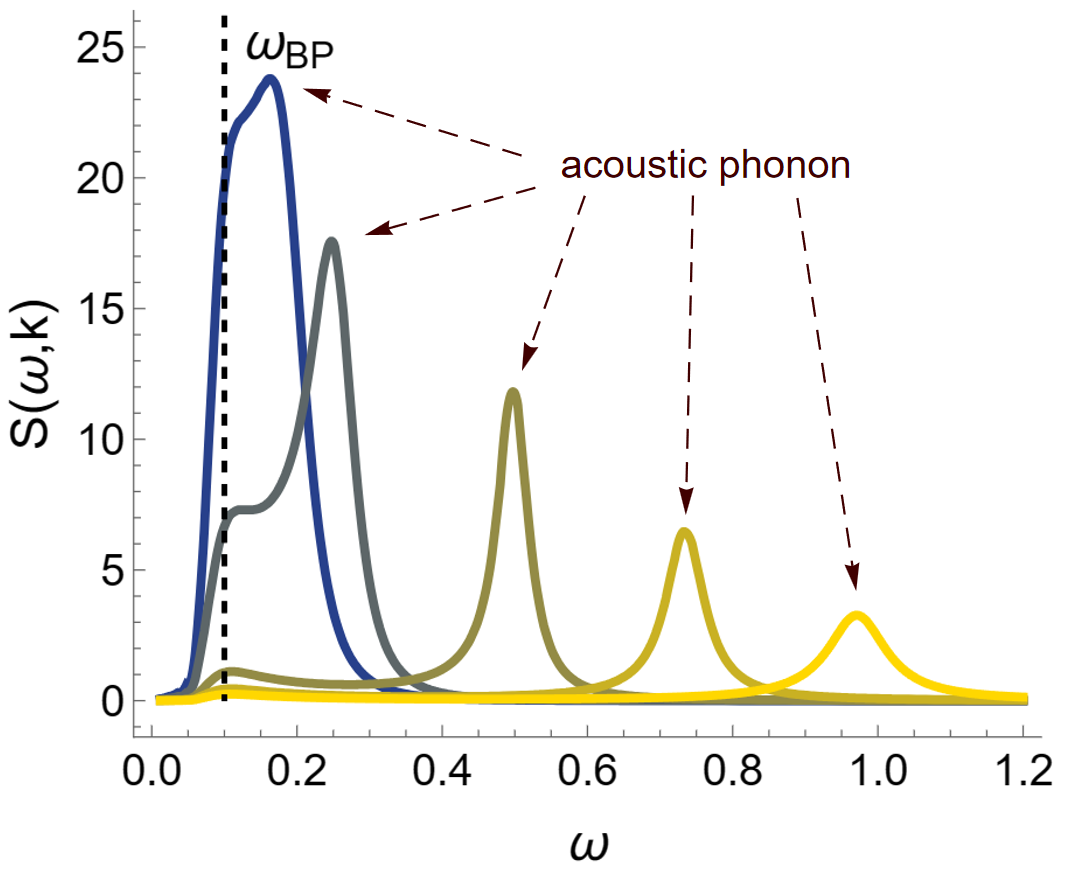}%
    \caption{\textbf{The BP mode in the dynamical structure factor.} The dynamical structure factor for $k=0.2,0.4,0.6,0.8$ from dark to light color. Other parameters are same as Fig.\ref{figdsf} (d). The vertical dashed line indicates the position of the dispersionless BP mode coinciding with $\omega_{\text{BP}}$. The peak corresponding to the acoustic phonon is also indicated.}
    \label{figs2}
\end{figure}

In order to confirm the existence and properties of the BP mode, in Fig.\ref{figs2}, we show several cuts of the dynamic structure factor at fixed wave-vector $k$ and varying frequency using the same parameters as in panel (d) of Fig.\ref{figdsf}. The existence of a dispersionaless mode, in addition to the propagating acoustic phonon, is clear from that representation. Let us also notice that the BP mode exists only for frequencies below the phonon frequency. Its flat dispersion ends on the linear dispersion of the acoustic phonon.

All the aforementioned properties of the flat BP mode are consistent with what observed in simulations by Hu and Tanaka \cite{Hu2022}. The existence of a flat mode, responsible for the BP anomaly, has also been experimentally observed, see for example \cite{Tomterud2023}. As a general remark, we highlight that from a macroscopic perspective (but certainly not from a microscopic point of view) the existence of such a flat mode is similar to the appearance of low-energy optical modes, that have been identified as the origin of the BP in several crystalline materials \cite{PhysRevB.99.024301,Baggioli_2020,doi:10.1021/acs.jpclett.2c01224}.

\subsection{Density of states and boson peak}
The most direct way to study the properties of BP is to investigate the reduced VDOS, $g(\omega)/\omega^{d-1}$, where $\omega^{d-1}$ is the scaling predicted by Debye law and $d$ the spatial dimension of system. 

\begin{figure}[htb]
    \centering
    \includegraphics[width=\linewidth]{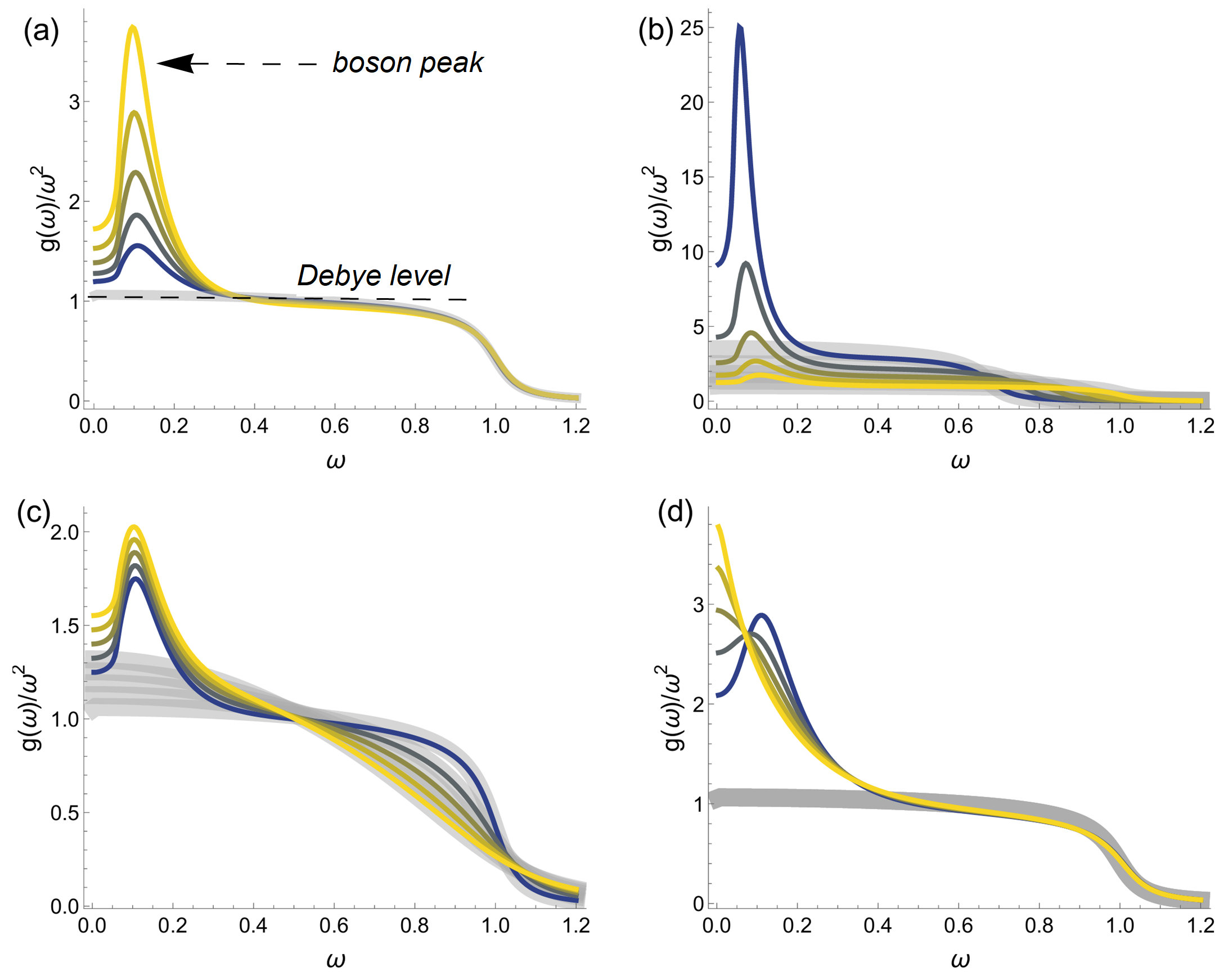}%
    \caption{\textbf{The boson peak.} \textbf{(a)} The Debye reduced VDOS for different values of the stringlet-phonon coupling $\mu=0.03, 0.05, 0.07, 0.09, 0.11$ (from blue to yellow). The parameters are fixed to $v=v_s=1$, $k_D=1$, $\lambda=5$ and $\gamma=\gamma_s=10^{-8}$. In the other panels, we dial respectively $v = 0.68 \to 1$ from blue to yellow (\textbf{b}),  $\gamma = 0.1 \to 0.5$ from blue to yellow (\textbf{c}), and $\gamma_s = 0.05 \to 0.25$ from blue to yellow (\textbf{d}). The other parameters involved are kept fixed. In all panels, gray lines are the results for $\mu=0$, displaying the case of the pure Debye model.}
    \label{figvdos1}
\end{figure}

In panel (a) of Fig.\ref{figvdos1}, we show the reduced VDOS as a function of the frequency for different values of the stringlet-phonon coupling $\mu$. The gray line shows the result of the pure Debye model with $\omega_D=1$. By increasing the coupling strength $\mu$, two phenomena are evident. First, the Debye level at low frequency increases. This is simply the consequence of the renormalization of the phonon speed of propagation, that gets smaller by enhancing the interactions with the stringlets acting as scatterers. Second, and most importantly, a clear BP excess anomaly emerges. The position of this anomaly, $\omega_{\text{BP}}$, coincides exactly with the energy of the flat mode reported in Fig.\ref{figdsf} and Fig.\ref{figs2}. The intensity of this anomaly becomes larger by increasing the coupling strength, while its position is only mildly dependent on it. The position is indeed mostly controlled by the average stringlet length (that is kept fixed in Fig.\ref{figvdos1}) and the phonon propagation speed (also fixed therein), consistent with the findings in \cite{jiang2023stringlet}.

The theoretical model at hand allows for a broad investigation of the BP properties as a function of the various parameters, that are analyzed in the other panels of Fig.\ref{figvdos1}, To reduce the number of parameters, the speed of stringlet propagation $v_s$ is always considered to be same as the speed of sound $v$.

As shown in panel (b) of Fig.\ref{figvdos1}, the intensity of the boson peak decreases by increasing $v$. We notice that this behavior is consistent with the trend of the experimental data, \textit{e.g.} \cite{experimentbpandspeed}. Moreover, $\omega_{\text{BP}}$ grows monotonically by increasing $v$ as the material becomes more rigid.

Finally, in panels (c) and (d) of Fig.\ref{figvdos1} we explore the properties of the BP by increasing the damping parameters $\gamma$, $\gamma_s$. Increasing the damping $\gamma$ makes the intensity of the BP grow, while its position is not strongly affected by it. More interesting is the result as a function of the stringlet damping $\gamma_s$. More precisely, $\gamma_s$ softens the BP energy scale. At a critical value of $\gamma_s$, the BP disappears into a $\omega = 0$ peak and the reduced DOS is monotonically decaying at low frequency. This trend is agreement with the results in \cite{jiang2023stringlet}, where this phenomenon has been studied in more detail.

\subsection{Sound attenuation and effective phonons speed}
\begin{figure}[htb]
    \centering
    \includegraphics[width=0.7\linewidth]{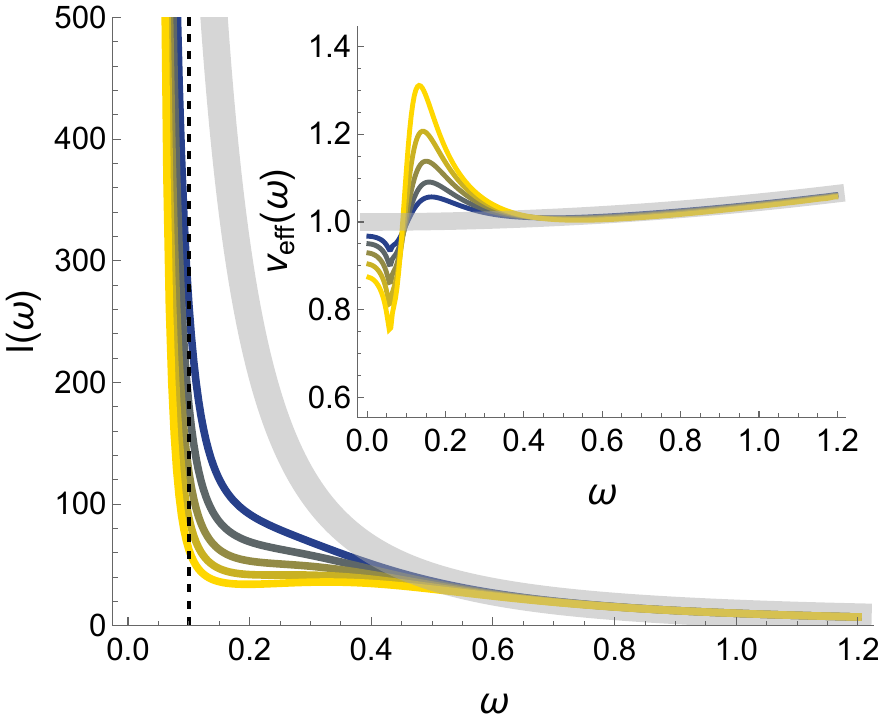}%
    \caption{\textbf{Phonon propagation in stringlet-land.} The sound attenuation length $l(\omega)$ and effective speed of propagation $v_{\text{eff}}(\omega)$ as a function of the frequency for increasing values of the stringlet-phonon coupling $\mu$. The gray lines correspond to the results for $\mu=0$. The other parameters involved here are the same as Fig.\ref{figvdos1} (the color scheme is also matched).}
    \label{figattenuation}
\end{figure}

The effective dispersion relation of the acoustic phonons can be obtained by considering the poles of the dressed propagator Eq.\eqref{g1},
\begin{equation}
    -\omega^2 +v^2 k^2-i \omega \gamma k^2- \mu^2 k^2 \tilde \Gamma(\omega)=0.
\end{equation}
We treat the wavevector $k$ as a complex number and we set the frequency $\omega$ to be a real number. Then, the equation above can be recasted as,
\begin{equation}
    k^2 \left(v^2-i \omega \gamma-\mu^2 \tilde \Gamma(\omega)\right)-\omega^2=0.
\end{equation}
Therefore, the effective dispersion relation is 
\begin{equation}
    k(\omega)=\dfrac{\omega}{\sqrt{v^2-i \omega \gamma-\mu^2 \tilde \Gamma(\omega)}}\equiv k_1(\omega)+i k_2(\omega).
\end{equation}
Following standard definitions, the effective speed of sound can be obtained as:
\begin{equation}
    v_{\text{eff}}(\omega)=
    \left(\frac{d k_1(\omega)}{d \omega}\right)^{-1},
\end{equation}
and the sound attenuation length is defined as the inverse of imaginary part of the wave-vector,
\begin{equation}
    l(\omega)=\frac{1}{k_2(\omega)}.
\end{equation}
In absence of stringlet-phonon interactions, $\mu=0$,
\begin{align}
    &k_1(\omega)=\frac{\omega \cos \left(\frac{1}{2} \arg \left(v^2-i \gamma  \omega\right)\right)}{\sqrt[4]{v^4+\gamma ^2 \omega^2}},\label{p1}\\
    &k_2(\omega)=-\frac{\omega \sin \left(\frac{1}{2} \arg \left(v^2-i \gamma  \omega\right)\right)}{\sqrt[4]{v^4+\gamma ^2 \omega^2}}.\label{p2}
\end{align}
In that case, $\gamma=0$ implies that $k_2(\omega)=0$ and $k_1(\omega)=\omega/v$ as expected. Two examples of $l(\omega)$ and $v_{\text{eff}}(\omega)$ following from Eqs.\eqref{p1}-\eqref{p2} are shown in Fig.\ref{figattenuation} in gray color.

In Fig.\ref{figattenuation} we show the sound attenuation length and the effective sound velocity as a function of the frequency by dialing the stringlet-phonon coupling $\mu$. The first visible feature is that the sound attenuation length diverges at $\omega=0$. This is simply the consequence of the absence of any scattering or damping mechanism that remains active in the limit of zero frequency. In other words, the imaginary part of the phonon propagator vanishes at $\omega=0$. A second property is that the larger $\mu$ the shorter attenuation length, that is compatible with $\mu$ determining the amount of interaction/scattering between stringlets and phonons. 

More in general, we observe that the stringlets renormalize strongly the phonon properties only in a frequency window around the BP energy. The sound attenuation length is strongly decreased slightly above the BP frequency, in the interval $\omega_{\text{BP}}<\omega<0.4$ in Fig.\ref{figattenuation}.

The effective phonon velocity is also strongly modified by the coupling with the stringlets. In particular, as evident from the inset in Fig.\ref{figattenuation}, the speed of sound is decreased slightly below the BP frequency and increased slightly above it.

Finally, for large frequencies, the interactions with the stringlets become subleading and both the sound attenuation and the effective speed approach their  $\mu=0$ values.

The picture just presented is consistent with the results of \cite{PhysRevB.101.174311}, proving that this phenomenology is quite insensitive to the precise form of the stringlet size distribution, aside from the information about the average length. 

\subsection{The length scale associated to the boson peak}
The idea of associating a length-scale to the BP is certainly not new and has been explored in various directions, \textit{e.g.}, \cite{PhysRevLett.68.974,PhysRevE.83.061508,HONG2011351,10.1063/1.2360275,MALINOVSKY1988111,MALINOVSKY1986757,https://doi.org/10.1002/pssb.2220640120}.

From a macroscopic perspective, in the spirit of heterogeneous elasticity theory \cite{doi:10.1142/9781800612587_0009}, the BP scale can be identified from the disordered distribution of the shear modulus. In other words, the length scale is the average size of the shear modulus fluctuations in the disordered solid. From a more microscopic perspective, the same length scale can be thought as arising from quasi-localized structures coexisting with phonons in amorphous materials \cite{10.1063/5.0069477}. In this context, the length scale corresponds to the average size of these ``defects'' that punctuate the otherwise homogeneous elastic medium. Interestingly, a direct quantitative relation between the scales defined using these two approaches has been confirmed in certain glass models \cite{PhysRevLett.127.215504}, providing a possible ``peace treaty'' between the two parts.

According to the stringlet theory of the boson peak \cite{PhysRevLett.127.215504}, the length-scale associated with the BP is the average stringlet size, determined by the stringlet length distribution $p(l)$. As shown by means of simulations in \cite{Hu2022}, and now proved theoretically (see Fig.\ref{figdsf}), stringlet-phonon interactions induce the emergence of a flat BP mode. We notice that, because of its non-dispersive character and its low-energy compared to that of acoustic phonons, it is conceivable to interpret such a mode as a quasi-localised low-energy excitation. We notice that the analysis in \cite{Hu2022} suggests that this BP mode is unrelated to the quadrupolar four-leaf defects discussed in the literature, that appear somehow at frequencies much below the BP scale, but this debate has not been settled yet \cite{10.1063/5.0147889}. 

Let us go back to the dressed phonon Green's function in Eq.\eqref{g1}. From there, we observe that the phonon speed is renormalized by the self-energy $\Gamma(\omega,k)$,
\begin{equation}
    \tilde v^2(k)=v^2-\dfrac{\mathrm{Re}[\Gamma(\omega (k),k)]}{k^2}.
\end{equation}
Here, we have used the dispersion relation $\omega(k)$ to make the renormalized speed a function of only the wave-vector $k$. At this point, it is straightforward to Fourier transform $\tilde v^2(k)$ back into real space and define a space-dependent phonon speed,
\begin{equation}
    \tilde v^2(x)=\mathcal{F}\left[\tilde v^2(k)\right].
\end{equation}
Microscopically, the heterogeneous nature of the phonon speed of propagation is just a direct consequence of the framework depicted in Fig.\ref{fig0}, in which the phonons propagate in a bath of vibrating stringlets with different sizes. After a coarse-graining procedure, that is technically realized via solving the Dyson's equations and integrating out the stringlet degrees of freedom, $\tilde v^2(x)$ is one of the macroscopic effects remaining.
 
\begin{figure}[htb]
    \centering
    \includegraphics[width=0.5\linewidth]{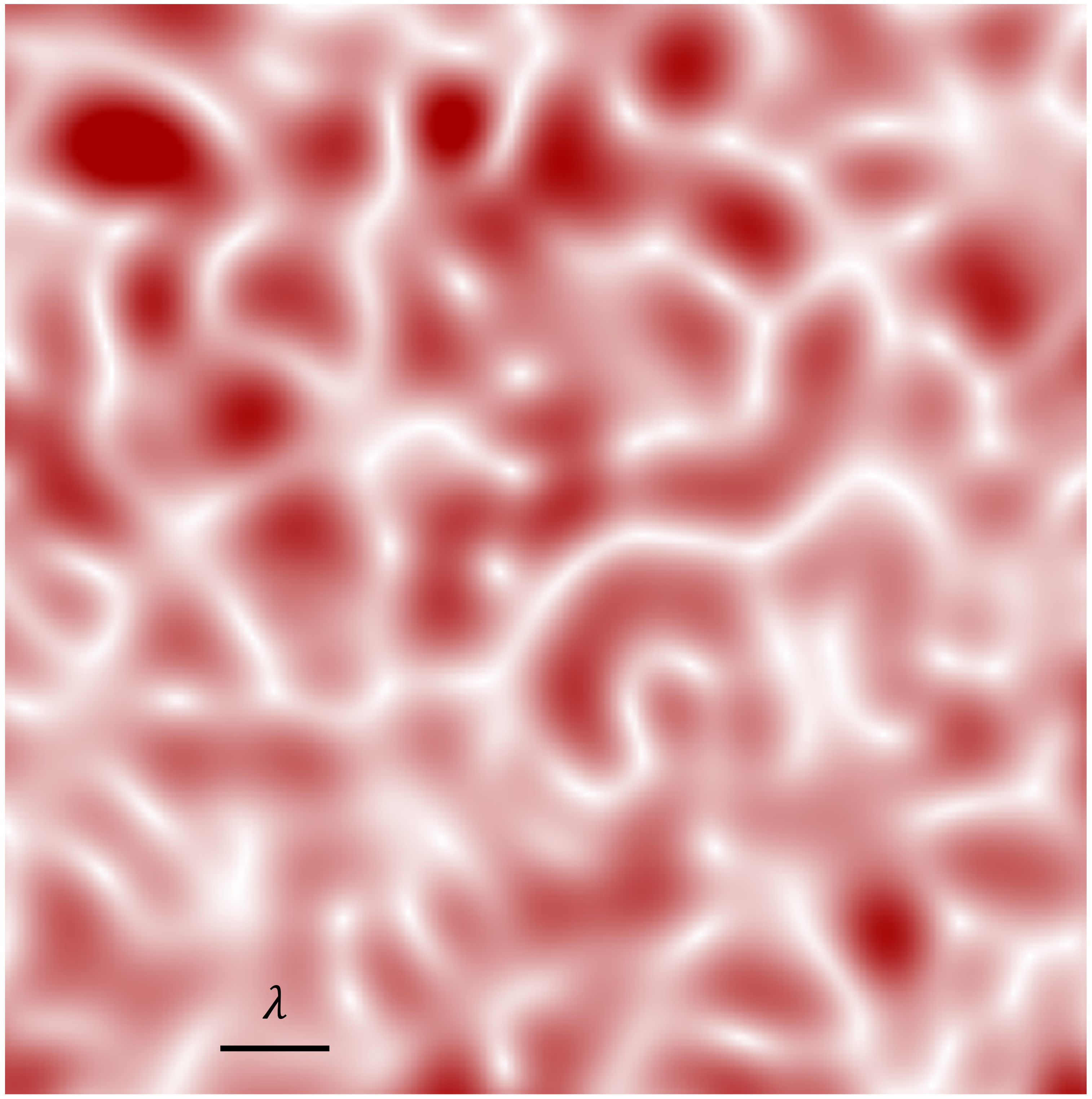}%
    \caption{\textbf{From stringlets to heterogeneous dynamics.} The heterogeneous phonon speed $\tilde v^2(x)$ in real space. The color map indicates its intensity from white to red color. The black line indicates the scale of $\lambda$ used to realize this figure.}
    \label{shearmodulusfluctuationremake}
\end{figure}

In Fig.\ref{shearmodulusfluctuationremake}, we provide a visual representation of this effect by using a benchmark value for the parameter $\lambda$ that characterizes the average stringlet length. Red color indicates spatial regions in which the phonon speed is large. The correlation between the average size of the fluctuations of $\tilde v^2(x)$ and $\lambda$ is evident, and perhaps not surprising. This simple analysis suggests that the ideas that the BP scale is set by the fluctuations of heterogeneous elasticity or by the average size of localized defects (whatever their microscopic origin is) are not incompatible. Indeed, in our view, they are just different approaches towards the same problem, one more macroscopic and effective, while the other more microscopic.

\section{Outlook and discussion}
In this work, we studied the dynamical and vibrational properties of an idealized amorphous solid modelled as an elastic medium punctuated by pinned elastic string-like obects (stringlets) with exponentially distributed length. The theoretical framework is a generalization of Lund's theory \cite{PhysRevB.101.174311} that uses an exponential stringlet size distribution verified by simulations. Despite the simplicity of the model, the results qualitatively agree with the observed BP properties and, together with \cite{jiang2023stringlet}, provide a theoretical ground for the recent simulation observations by Hu and Tanaka \cite{Hu2022,PhysRevResearch.5.023055}. 

Several open questions remain. In particular, at present, the emergent nature of the stringlets remains unclear. Stringlets are certainly not fundamental vibrational excitations but rather an emergent collective phenomenon that arises probably because of interactions and anharmonicity. The fundamental origin of the stringlets, and the statistical mechanics reasoning behind their appearance itself, are still matter of investigation. Moreover, from a theoretical point of view, it is not obvious to understand why in a 3D amorphous solid, the fundamental objects behind the BP anomaly should be of 1D nature. At this moment, this seems to be only a simplifying working assumption that needs further corroboration.

Finally, the emergence of a BP flat mode is a very appealing feature. First, this is in common among various theoretical approaches including heterogeneous elasticity theory and quasi-localized modes. Second, this feature is observed not only in simulations \cite{Hu2022} but even in experiments \cite{Tomterud2023}. Third, from a macroscopic point of view, this feature is strikingly similar to the existence of low-lying optical modes that have been ascribed as the origin of the BP anomaly in crystalline materials \cite{PhysRevB.99.024301,Baggioli_2020,doi:10.1021/acs.jpclett.2c01224}, hinting toward a possible macroscopic (but not microscopic) universal framework. Along these lines, stringlets have been indeed already discussed as the responsible for the BP anomaly in heated crystals by Douglas and collaborators \cite{C2SM27533C,C2SM26789F}. These analogies certainly deserve more attention in the near future.

\section*{Acknowledgments}
We would like to thank Jack Douglas for many useful conversations on the topic and collaboration on related topics. We would like to thank Jie Zhang, Alessio Zaccone and Massimo Pica Ciamarra for many discussions about the boson peak. We thank Lijin Wang for useful comments. We thank Jia-Lin Wu for useful comments and suggestion on this manuscript and for future research.
We acknowledge the support of the Shanghai Municipal Science and Technology Major Project (Grant No.2019SHZDZX01). MB acknowledges the sponsorship from the Yangyang Development Fund.

\appendix
\section{Details of the theoretical models and mathematical derivations}\label{details}
In this Appendix, we provide further details and mathematical derivations regarding the formulae presented in the main text.

We start by discussing the bare phonon Green's function in Eq.\eqref{bareph}. Newton's law expressed in terms of the displacement vector $u_i$ is given by
\begin{equation}\label{newton}
    \rho  \partial_t^2 u_i=\nabla_j \sigma_{ij}+f_i^{\text{ext}},
\end{equation}
where $\sigma_{ij}$ is the stress tensor, $f_i^{\text{ext}}$ the total external force and $\rho$ the mass density. The stress tensor can be decomposed in an elastic part and a viscous part,
\begin{equation}
\sigma_{ij}=\sigma_{ij}^{\text{el}}+\sigma_{ij}^{\text{visc}}=C_{ijkl}\epsilon_{kl}+\eta_{ijkl} \partial_t \epsilon_{kl},\label{lala}
\end{equation}
where $\epsilon_{ij}=1/2\left(\nabla_i u_j+\nabla_j u_i\right)$ is the linear strain tensor. By plugging Eq.\eqref{lala} into Newton's law, and assuming that the external forces vanish, one obtains
\begin{equation}\label{sisi}
     \rho  \partial_t^2 u_i=C_{ijkl} \nabla_j \epsilon_{kl}+\eta_{ijkl} \partial_t \nabla_j \epsilon_{kl}.
\end{equation}
The above equation can be further decomposed into transverse and longitudinal components as done in standard viscoelasticity theory. By dropping these complications and focusing on only one polarization, the Green's function for the corresponding differential equation can be easily derived in Fourier space,
\begin{equation}
    \left(-\omega^2+v^2 k^2-i \omega k^2 \gamma\right) \mathcal{G}_0(\omega,k)=1,
\end{equation}
that yields directly to the bare phonon Green's function presented in the main text in Eq.\eqref{bareph}. Notice how the vanishing of the imaginary part in the limit $k \rightarrow 0$ is a direct consequence of translational invariance and the Goldstone nature of acoustic phonons.

One can then introduce the interaction between the phonon modes and the stringlets that will lead to a non-trivial self-energy correction $\Gamma(\omega,k)$. The various contributions in arbitrary Feynman diagrams can be re summed in closed form using the so-called Feynman-Dyson perturbation theory \cite{coleman2015introduction} that ultimately leads to the Dyson equation, Eq.\eqref{dys} in the main text. Eq.\eqref{g1} follows directly from this standard many-body physics procedure \cite{coleman2015introduction}.

Finally, let us provide more details about the derivation of the self-energy $\Gamma(\omega,k)$ in Eq.\eqref{selfie}. The equation presented in the main text is a simplified version of a more general result obtained by Lund and collaborators. In particular, the interested Reader can find an exhaustive discussion in Refs.~\cite{PhysRevB.72.174110,PhysRevB.72.174111}. Below, we briefly reproduce the main results. We will try for consistency to be as close as possible in notations to the original works and we will describe step by step the mapping to the notations that we have used in the main text.

We start by considering a stringlet segment $\textbf{X}(s,t)$ where $s$ is the coordinate along the one-dimensional stringlet and $t$ is the time coordinate. In the approximation of neglecting friction (and the corresponding damping coefficient), the equation of motion for $\textbf{X}(s,t)$ is given by
\begin{equation}\label{ppp}
    m \ddot{X}_k(s,t)- \zeta X_k''(s,t)=F_k,
\end{equation}
where $k$ is just the index of the vector $\textbf{X}$, dot stands for time derivative and $'$ for derivative with respect to the line coordinate $s$. Here, $m,\zeta$ are parameters that map to $v_s$ in Eq.\eqref{vuvu} in the main text, $v_s^2=\zeta/m$. $F_k$ represents the force on the stringlet. Following the original idea of considering stringlets as dislocation-lines, we model $F_k$ using the Peach-Koehler force \cite{PhysRev.80.436},
\begin{equation}
    F_k=  N_{kjp} \nabla_j u_p(\textbf{X}_0,t),
\end{equation}
where $N_{kjp}$ is a dimensionful tensor and $\textbf{X}_0$ the equilibrium position of the stringlet. This force is the result of the stress induced by a sound wave hitting on the stringlet. The backreaction of the vibrating stringlets on the propagation of sound waves can be then modelled as a source $s_i$ in the equation for the phonons that, written in terms of the velocities $v_i=\dot u_i$, reads
\begin{equation}\label{nnn}
    \rho \ddot{v}_i(\textbf{x},t)- C_{ijkl} \frac{\partial^2}{\partial x_j \partial x_l} v_k(\textbf{x},t)=s_i(\textbf{x},t),
\end{equation}
where again for simplicity we have neglected dissipative corrections (corresponding to the $\eta_{ijkl}$ tensor in Eq.\eqref{sisi}). The source term $s_i$ obeys the following relation,
\begin{equation}\label{sss}
    s_i(t)= M_{ijk}\int_L ds \dot{X}_j(s,t) \nabla_k \delta\left(\textbf{x}-\textbf{X}_0\right),
\end{equation}
where $M_{ijk}$ is a dimensionful tensor whose form can be explicitly derived in the case of dislocation lines, and $L$ indicates integration over the full stringlet segment.

By combining the solution of Eq.\eqref{ppp} with the definition of the source in Eq.\eqref{sss}, $s_i$ can be written in terms of a potential $V_{ik}$ \cite{PhysRevB.101.174311}, $s_i=V_{ik} v_k$, that is given by,
\begin{equation}\label{popo}
    V_{ik}= A_{ikjl} \,l\,\mathcal{G}_s(\omega,l) \nabla_j \delta\left(\textbf{x}-\textbf{X}_0\right) \nabla_l|_{\textbf{x}=\textbf{X}_0}
\end{equation}
where $A_{ikjl}$ is a dimensionful tensor, whose complete structure can be found in Refs.~\cite{PhysRevB.72.174110,PhysRevB.72.174111}, and $\mathcal{G}_s(\omega,l)$ is the bare stringlet Green's function defined in the main text, Eq.\eqref{vuvu}. We notice that this result holds for one single stringlet. If we take a distribution of them, defined by their length distribution function $p(l)$, one then has to integrate over this distribution to obtain the final result for the average Green's function. We will do that below. By using Eq.\eqref{popo} into Eq.\eqref{nnn}, and going to Fourier space, we get
\begin{equation}
   - \rho \,\omega^2 v_i- C_{ijkl} \,k_j k_l \,v_k= A_{ikjl}\,k_j\,k_l\, v_k\,\int l\,p(l)  \,\mathcal{G}_s(\omega) dl.
\end{equation}
Finally, by assuming isotropy and considering only one polarization, the above equation can be reduced in the simpler form,
\begin{equation}
   - \rho \,\omega^2 v_i- G \,k^2 \,v_i= \mu^2\,k^2\, v_i\,\int l\,p(l)  \,\mathcal{G}_s(\omega) dl,
\end{equation}
that leads to the identification
\begin{equation}\label{fifi}
    \Gamma(\omega,k)=\mu^2\,k^2\,\int l\,p(l)  \,\mathcal{G}_s(\omega) dl
\end{equation}
presented in the main text as Eq.\eqref{selfie}. We notice that $\Gamma(\omega,k)$ in our notations correspond to $f(\omega)k^2$ in the notations of Ref. \cite{PhysRevB.101.174311}. One can immediately verify that the final result therein, Eq.(A14), coincides exactly with our expression in Eq.\eqref{fifi} when the second order correction in perturbation theory is turned off, $\alpha \rightarrow 0$ in the language of \cite{PhysRevB.101.174311}, and when $f_0$ is identified with our $\mu^2/v^2$.

In summary, modulo re-definitions, the formulas used in the main text coincide exactly with those derived in \cite{PhysRevB.101.174311}, with the difference that in our case $p(l)$ is assumed to be exponential, as corroborated in simulations, and not treated as a phenomenological fitting function to be extracted from the experimental data.

\vspace{0.3cm}

\def\url#1{}
\providecommand{\newblock}{}

\end{document}